\documentclass[12pt]{article}
\usepackage{amssymb}
\renewcommand{\theequation}{\arabic{section}.\arabic{equation}}

\begin{document}



\def\a{\bar{\alpha}}
\def\b{\beta}
\def\d{\delta}
\def\e{\epsilon}
\def\g{\gamma}
\def\h{\mathfrak{h}}
\def\k{\kappa}
\def\l{\lambda}
\def\o{\omega}
\def\p{\wp}
\def\t{\theta}
\def\s{\sigma}
\def\z{\zeta}
\def\x{\xi}
\def\w{\sqrt{-1}g}
\def\A{{\cal{A}}}
\def\B{{\cal{B}}}
\def\C{{\cal{C}}}
\def\D{{\cal{D}}}
\def\G{\Gamma}
\def\H{\cal{H}}
\def\O{\Omega}
\def\L{\Lambda}
\def\f{E_{\tau,\eta}(sl_2)}
\def\E{E_{\tau,\eta}(sl_n)}
\def\r{E_{\tau,\eta}(sl_{n-1})}
\def\Zb{\mathbb{Z}}
\def\Cb{\mathbb{C}}
\def\R{\mathbb{R}}

\def\beq{\begin{equation}}
\def\eeq{\end{equation}}
\def\bea{\begin{eqnarray}}
\def\eea{\end{eqnarray}}
\def\ba{\begin{array}}
\def\ea{\end{array}}
\def\no{\nonumber}
\def\le{\langle}
\def\re{\rangle}
\def\lt{\left}
\def\rt{\right}

\newtheorem{Theorem}{Theorem}
\newtheorem{Definition}{Definition}
\newtheorem{Proposition}{Proposition}
\newtheorem{Lemma}{Lemma}
\newtheorem{Corollary}{Corollary}
\newcommand{\proof}[1]{{\bf Proof. }
        #1\begin{flushright}$\Box$\end{flushright}}

\baselineskip=20pt

\newfont{\elevenmib}{cmmib10 scaled\magstep1}
\newcommand{\preprint}{
   \begin{flushleft}
     \elevenmib Yukawa\, Institute\, Kyoto\\
   \end{flushleft}\vspace{-1.3cm}
   \begin{flushright}\normalsize  \sf
     YITP-03-13\\
     {\tt hep-th/0309194} \\ September 2003
   \end{flushright}}
\newcommand{\Title}[1]{{\baselineskip=26pt
   \begin{center} \Large \bf #1 \\ \ \\ \end{center}}}
\newcommand{\Author}{\begin{center}
   \large \bf B.\,Y.~Hou${}^a$, ~R.~Sasaki${}^b$~ and~W.-L. Yang${}^{a,b}$ \end{center}}
\newcommand{\Address}{\begin{center}
     $^a$ Institute of Modern Physics, Northwest University\\
     Xian 710069, P.R. China\\
     ~~\\
     ${}^b$ Yukawa Institute for Theoretical Physics,\\
     Kyoto University, Kyoto 606-8502, Japan
   \end{center}}
\newcommand{\Accepted}[1]{\begin{center}
   {\large \sf #1}\\ \vspace{1mm}{\small \sf Accepted for Publication}
   \end{center}}

\preprint
\thispagestyle{empty}
\bigskip\bigskip\bigskip

\Title{Eigenvalues of  Ruijsenaars-Schneider models associated
with $A_{n-1}$ root system in  Bethe ansatz formalism } \Author

\Address
\vspace{1cm}

\begin{abstract}
Ruijsenaars-Schneider models associated with $A_{n-1}$ root system
with a discrete {\it coupling constant\/} are studied. The
eigenvalues of the Hamiltonian are given in terms of the Bethe
ansatz formulas. Taking the ``non-relativistic" limit, we obtain
the spectrum of the corresponding Calogero-Moser systems in the
{\it third formulas\/} of Felder et al \cite{Fel95}.

\vspace{1truecm} \noindent {\it PACS:} 03.65.Fd; 05.30.-d

\noindent {\it Keywords}: Integrable model; Elliptic quantum
group; Bethe ansatz;  Ruijsenaars-Schneider model; Calogero-Moser
model.
\end{abstract}
\newpage
\section{Introduction}
\label{intro}
\setcounter{equation}{0}

Ruijsenaars-Schneider (RS) models \cite{Rui86,Rui87} are
integrable generalization of Calogero-Moser (CM) models
\cite{Cal75, Mos75} at both classical and quantum levels. The
integrability of classical RS models associated with various root
systems were studied by Lax pair representation  for $A_{n-1}$
\cite{Rui86}, for $D_n$ \cite{Hou01}, for  $C_n$ and $BC_n$ with
one  coupling constant \cite{Hou001}. The commuting conserved
quantities for quantum RS models were discussed in
Refs\cite{Rui87,Die94,Kom98,Has99}.

It is well known that the Hamiltonian of  RS model with degenerate
potentials (trigonometric and rational ones) is one of the
commuting families of  Macdonald operators \cite{Mac95}, which are
also called {\it Ruijsenaars-Macdonald operators\/}. The
eigenfunctions of the {\it degenerate\/} Hamiltonian are given by
so-called {\it Macdonald polynomials\/} \cite{Mac95}. However, an
analogous  construction for elliptic generalization of {\it
Macdonald polynomials\/} is still an open problem.

Bethe ansatz method  has  proved to be the most powerful and
(probably) unified method to construct the common eigenvectors of
commuting families of operators (usually called {\it transfer
matrices\/}) in two-dimensional trigonometric and rational
integrable models \cite{Fad79,Tha82,Bab82, Sch83,Kor93}. Recently,
after a definition of elliptic quantum groups
$E_{\tau,\eta}(\mathfrak{g})$ associated with any simple classical
Lie algebra $\mathfrak{g}$ was given \cite{Fel94},  the algebraic
Bethe ansatz method has been successfully extended. The method for
construction of the eigenvectors of the transfer matrices
associated with the module over $E_{\tau,\eta}(sl_2)$ \cite{Fel96}
is now generalized to apply for those associated with the module
over $E_{\tau,\eta}(sl_n)$ with generic $n$ \cite{Bil98,Hou03}.

In particular, the elliptic {\it Ruijsenaars-Macdonald operator}
(\ref{Rui-elliptic}) with a discrete  {\it coupling constant\/}
$\g=\w l$ ($l$ being a non-negative integer) associated with
$A_{n-1}$ root system can be rewritten as the {\it transfer
matrices\/} associated with the symmetric $n\times l$ tensor
product evaluation $E_{\tau,\eta}(sl_n)$-module \cite{Fel961}.
This enables us to obtain the eigenvalues of the Hamiltonian of
elliptic RS models with the discrete {\it coupling constant\/}
$\g=\w l$ associated with $A_{n-1}$ root system by the algebraic
Bethe ansatz for elliptic quantum group $E_{\tau,\eta}(sl_n)$.

The paper is organized as follows. In section 2, we give a brief
review of the algebraic Bethe ansatz for elliptic quantum group
$E_{\tau,\eta}(sl_n)$ developed in Ref.\cite{Hou03}. In section 3,
choosing the special $\E$-module $W=V_{\L^{(nl)}}(0)$, we give the
eigenvalues of elliptic RS model with the discrete {\it coupling
constant\/} $\g=\w l$ and  the associated Bethe ansatz equations.
Taking the degenerate (trigonometric and rational) limit, we
obtain the eigenvalues of Hamiltonian of the degenerate RS models
associated with $A_{n-1}$ root system. In section 4, taking the
``non-relativistic" limit \cite{Bra97}, we obtain the eigenvalues
of elliptic, trigonometric and rational types of CM models
associated with $A_{n-1}$ root system with an integer {\it
coupling constant\/} $\g=l+1$ in the Bethe ansatz formulas or the
{\it third formulas\/} in the sense of Felder et al \cite{Fel95}.

\section{Algebraic Bethe ansatz for  elliptic quantum group $\E$}
\label{ABAE} \setcounter{equation}{0}
\subsection{Elliptic quantum group associated  with $A_{n-1}$}

We first review the definition of the elliptic quantum group $\E$
associated with  $A_{n-1}$ \cite{Fel94}. Let
$\lt\{\e_{i}~|~i=1,2,\cdots,n\rt\}$ be the orthonormal basis of
the vector space $\Cb^n$ such that $\langle\e_i,~\e_j
\rangle=\d_{ij}$. The $A_{n-1}$ simple roots are
$\lt\{\a_{i}=\e_i-\e_{i+1}~|~i=1,\cdots,n-1\rt\}$ and the
fundamental weights $\lt\{\L_i~|~i=1,\cdots,n-1\rt\}$ satisfying
$\langle\L_i,~\a_j\rangle=\d_{ij}$ are given by \bea
\L_i=\sum_{k=1}^{i}\e_k-\frac{i}{n}\sum_{k=1}^{n}\e_k. \no\eea Set
\bea \hat{i}=\e_i-\overline{\e},~~\overline{\e}=
\frac{1}{n}\sum_{k=1}^{n}\e_k,~~i=1,\cdots,n,~~{\rm
then}~~\sum_{i=1}^n\hat{i}=0. \label{Vectors} \eea For each
dominant weight $\L=\sum_{i=1}^{n-1}a_i\L_{i}~,~~a_{i}\in \Zb^+$,
there exists an irreducible highest weight finite-dimensional
representation $V_{\L}$ of $A_{n-1}$ with the highest vector $
|\L\rangle$. For example the fundamental vector representation is
$V_{\L_1}$. In this paper, we  consider only the symmetric
tensor-product representation of
$\stackrel{l}{\overbrace{V_{\L_1}\otimes V_{\L_1}\cdots \otimes
V_{\L_1}}}$ (or, the {\it higher spin-$l$ representation} of
$A_{n-1}$), namely, the one parameter series of highest weight
representations $V_{\L^{(l)}}$, with \bea
\L^{(l)}=l\L_1,~~l\in\Zb~~{\rm and~~} l>0\label{Weight}.\eea This
corresponds to the Young diagram
$\stackrel{l}{\overbrace{\Box\!\Box\!\Box\cdots\Box}}$.

Let $\h$ be the Cartan subalgebra of $A_{n-1}$ and $\h^{*}$ be its
dual. A finite dimensional diagonalisable  $\h$-module is a
complex finite dimensional vector space $W$ with a weight
decomposition $W=\oplus_{\mu\in \h^*}W[\mu]$, so that $\h$ acts on
$W[\mu]$ by $x~v=\mu(x)~v$, $(x\in \h,~~v\in~W[\mu])$. For
example, the fundamental vector representation $V_{\L_1}=\Cb^n$,
the non-zero weight spaces $W[\hat{i}]=\Cb \e_i,~i=1,\cdots,n$.

Let us fix $\tau$ such that $Im(\tau)>0$ and a generic complex
number $\eta$. For convenience, we introduce another parameter
$w=n\eta$ related to $\eta$.  Let us introduce  the following elliptic
functions \bea
&&\t\lt[\begin{array}{c} a\\b
\end{array}\rt](u,\tau)=\sum_{m=-\infty}^{\infty}
\exp\lt\{\sqrt{-1}\pi\lt[(m+a)^2\tau+2(m+a)(u+b)\rt]\rt\},\label{Theta}\\
&& \s(u)=\t\lt[\begin{array}{c}\frac{1}{2}\\[2pt]\frac{1}{2}
\end{array}\rt](u,\tau),~~~\z(u)=\partial_{u}\lt\{\ln
\s(u)\rt\},~~\p(u)=-\partial_{u}\lt\{\z(u)\rt\}. \label{Function}
\eea These functions have the following properties \bea
&&\s(u)=0+u\s'(0)+\frac{u^3}{6}\s'''(0)+\cdots,~~{\rm when}~
u\rightarrow 0, \label{P1} \\
&&\s(-u)=-\s(u),~~~\z(-u)=-\z(u),~~~\p(-u)=\p(u),\label{P2}\eea
where $\s'(0)=\partial_{u}\lt\{\s(u)\rt\}|_{u=0}$ and
$\s'''(0)=\partial^3_{u}\lt\{\s(u)\rt\}|_{u=0}$.

For a generic $\l\in \Cb^n$, define \bea
\l_i=\langle\l,\e_i\rangle,
~~\l_{ij}=\l_i-\l_j=\langle\l,\e_i-\e_j\rangle,~~i,j=1,\cdots,n.
\label{Def1}\eea Let $R(z,\l)\in End(\Cb^n\otimes\Cb^n)$ be the
R-matrix given by \bea
&&R(z,\l)=\sum_{i=1}^{n}R^{ii}_{ii}(z,\l)E_{ii}\otimes E_{ii}
+\sum_{i\ne j}\lt\{R^{ij}_{ij}(z,\l)E_{ii}\otimes E_{jj}+
R^{ji}_{ij}(z,\l)E_{ji}\otimes E_{ij}\rt\},\no\\
\label{R-matrix} \eea in which $E_{ij}$ is the matrix with
elements $(E_{ij})^l_k=\d_{jk}\d_{il}$. The coefficient functions
are  \bea &&R^{ii}_{ii}(z,\l)=1,~~
R^{ij}_{ij}(z,\l)=\frac{\s(z)\s(\l_{ij}+w)}
{\s(z+w)\s(\l_{ij})},\label{Elements1}\\
&& R^{ji}_{ij}(z,\l)=\frac{\s(w)\s(z+\l_{ij})}
{\s(z+w)\s(\l_{ij})},\label{Elements2}\eea  and  $\l_{ij}$ is
defined in (\ref{Def1}). The R-matrix satisfies the dynamical
(modified) quantum Yang-Baxter equation \bea
&&R_{12}(z_1-z_2,\l-wh^{(3)})R_{13}(z_1-z_3,\l)
R_{23}(z_2-z_3,\l-wh^{(1)})\no\\
&&~~~~=R_{23}(z_2-z_3,\l)R_{13}(z_1-z_3,\l-wh^{(2)})R_{12}(z_1-z_2,\l),
\label{MYBE}\eea with the initial condition \bea
R^{kl}_{ij}(0,\l)=\d^l_i~\d^k_j.\label{Initial}\eea We adopt the
notation: $R_{12}(z,\l-wh^{(3)})$ acts on a tensor $v_1\otimes v_2
\otimes v_3$ as $R(z;\l-w\mu)\otimes Id$ if $v_3\in W[\mu]$.

A representation of the elliptic quantum group $\E$ (an
$\E$-module) is  by definition a pair ($W,L$) where $W$ is a
diagonalisable $\h$-module and $L(z,\l)$ is a meromorphic function
of $\l$ and the spectral parameter $z\in\Cb$,  with values in
$End_{\h}(\Cb^n\otimes W)$ (the endomorphisms commuting with the
action of $\h$). It obeys the so-called $``RLL"$ relation \bea
&&R_{12}(z_1-z_2,\l-wh^{(3)})L_{13}(z_1,\l)
L_{23}(z_2,\l-wh^{(1)})\no\\
&&~~~~=L_{23}(z_2,\l)L_{13}(z_1,\l-wh^{(2)})R_{12}(z_1-z_2,\l),\label{Exchange}
\eea where the first and second space are auxiliary  spaces
($\Cb^n$) and the third space plays the role of the quantum space
($W$). The total weight conservation condition for the
$L$-operator reads \bea [h^{(1)}+h^{(3)},~L_{13}(z,\l)]=0.\no\eea
In terms of the elements of the $L$-operator defined by \bea
L(z,\l)\lt(\e_i\otimes v\rt)=\sum_{j=1}^n \e_j\otimes
L^j_i(z,\l)v,~~v\in W,\eea the above condition can be expressed
equivalently as \bea
f(h)L_i^j(z,\l)=L^j_i(z,\l)f(h+\hat{i}-\hat{j}),\label{Charge}\eea
in which $f(h)$ is any meromorphic function of $h$ and $h$
measures the weight of the quantum space ($W$).

\subsection{Modules over $\E$ and the associated operator algebra}
The basic example of an $\E$-module is $(\Cb^n, L)$ with
$L(z,\l)=R(z-z_1,\l)$, which is called  the fundamental vector
representation $V_{\L_1}(z_1)$ with the evaluation point $z_1$. It
is obvious that $``RLL"$ relation is satisfied as a consequence of
the dynamical Yang-Baxter equation (\ref{MYBE}). Other modules can
be obtained by taking tensor products: if $(W_1,L^{(1)})$ and
$(W_2,L^{(2)})$ are $\E$-modules, where $L^{(j)}$ acts on
($\Cb^n\otimes W_j$), then also $(W_1\otimes W_2, L)$ with \bea
L(z,\l)=L^{(1)}(z,\l-wh^{(2)})L^{(2)}(z,\l)~~{\rm acting~ on}
~\Cb^n\otimes W_1\otimes W_2.\label{Fusion}\eea

An $\E$-submodule of an $\E$-module $(W,L)$ is a pair $(W_1,L_1)$
where $W_1$ is an $\h$-submodule of $W$ such that $\Cb^n\otimes
W_1$ is invariant under the action of all the $L(z,\l)$, and
$L_1(z,\l)$ is the restriction to this invariant subspace. Namely,
the $\E$-submodules are $\E$-modules.

Using the fusion rule of $\E$ (\ref{Fusion}) one can construct the
symmetric $\E$-submodule of $l$-tensors of fundamental vector
representations: \bea V_{\L^{(l)}}(z_1)=~symmetric~
subspace~of~V_{\L_1}(z_1)\otimes
V_{\L_1}(z_1-w)\otimes\cdots\otimes V_{\L_1}(z_1-(l-1)w),\no \eea
where $\L^{(l)}$ is defined by (\ref{Weight}). We call such an
$\E$-module  the {\it higher spin-$l$ representation} with the
evaluation point $z_1$. These series of representations in the
case of $\Zb_n$ Sklyanin algebra have been studied in \cite{Skl83}
for $n=2$ case and in \cite{Hou97,Has97} for generic $n$ case.

For any $\E$-module, as in \cite{Fel96} one can define an
associated operator algebra of difference operators on the space
$Fun(W)$ of meromorphic functions of $\l$ with values in $W$. The
algebra is  generated by $h$ and the operators $\tilde{L}(z)\in
End(\Cb^n\otimes Fun(W))$ acting as\bea \tilde{L}(z)(\e_i\otimes
f)(\l)=\sum_{j=1}^{n}\e_j\otimes
L_i^{j}(z,\l)f(\l-w\hat{i}).\label{Definition2}\eea One can derive
the following exchange relation of the difference operator
$\tilde{L}(z)$ from the $``RLL"$ relation (\ref{Exchange}), the
weight conservation condition $L^j_i(z,\l)$ (\ref{Charge}) and the
fact $[h^{(1)}+h^{(2)},~R_{12}(z,\l)]=0$, \bea
&&R_{12}(z_1-z_2,\l-wh)\tilde{L}_{13}(z_1) \tilde{L}_{23}(z_2)
=\tilde{L}_{23}(z_2)\tilde{L}_{13}(z_1)R_{12}(z_1-z_2,\l),\label{Exchange1}\\
&&f(h)\tilde{L}_i^j(z)=\tilde{
L}^j_i(z)f(h+\hat{i}-\hat{j}),\label{Exchange2} \eea where $f(h)$
is any meromorphic function of $h$. Let $W=
V_{\L^{(l_1)}}(z_1)\otimes V_{\L^{(l_2)}}(z_2)\otimes
\cdots\otimes V_{\L^{(l_m)}}(z_m)$ and
$\L=\L^{(l_1)}+\cdots+\L^{(l_m)}$, then $W[\L]=\Cb |\L\rangle$
with $|\L\rangle=|\L^{(l_1)}\rangle\otimes \cdots
\otimes|\L^{(l_m)}\rangle$.
\begin{Theorem} \cite{Hou03} With
generic evaluation points $\lt\{z_i\rt\}$, $W$ is an irreducible
highest weight $\E$-module and  the vector $|\L\rangle$, viewed as
a constant function in $Fun(W)$, obeys the following highest
weight conditions: \bea
&&\tilde{L}^1_1(z)|\L\rangle=A(z,\l)|\L\rangle,~~\tilde{L}^i_1(z)|\L
\rangle=0,~~i=2,\cdots,n,\no\\
&&\tilde{L}^{i}_{j}(z)|\L\rangle=\d^{i}_{j}
D_i(z,\l)|\L\rangle,~~i,j=2,\cdots,n,~~~
f(h)|\L\rangle=f(N\hat{1})|\L\rangle.\no\eea The highest weight
functions read \bea A(z,\l)=1,~~
D_i(z,\l)=\lt\{\prod_{k=1}^{m}\frac{\s(z-p_k)}{\s(z-q_k)}\rt\}
\frac{\s(\l_{i1}+Nw)}{\s(\l_{i1})},~~i=2,\cdots,n,\label{Func}\eea
where \bea p_k=z_k,~~q_k=z_k-l_kw,~~~
N=\sum_{k=1}^{m}l_k,~~k=1,\cdots,m.\label{number}\eea
\end{Theorem}

\vspace{0.6truecm}

\noindent The {\it transfer matrices} associated with  an
$\E$-module $(W, L)$ \cite{Fel96} are  difference operators acting
on the space $Fun(W)[0]$ of meromorphic  functions of $\l$ with
values in the zero-weight space of $W$. They are defined by \bea
T(u)f(\l)=\sum_{i=1}^{n}\tilde{L}^i_i(u)f(\l)
=\sum_{i=1}^{n}L^{i}_{i}(u,\l)f(\l-w\hat{i}). \label{transfer}\eea
The exchange relations of $\tilde{L}$-operators (\ref{Exchange1})
and (\ref{Exchange2}) imply that, for any $\E$-module, the above
transfer matrices preserve the space ${\H}=Fun(W)[0]$. Moreover,
they commute pairwise on ${\H}$: $\lt[T(u)~,~T(v)\rt]|_{{\H}}=0$.

\subsection{Algebraic Bethe ansatz for $\E$}
We fix a highest weight $\E$-module $W$ of weight $\L$, the
functions $A(z,\l), D_i(z,\l)$  (\ref{Func}), with the highest
weight vector $|\L\rangle$. We assume that
$N=\sum_{k=1}^{m}l_k=n\times l$ with $l$ being an integer, so that
the zero-weight space $W[0]$ can be non-trivial and  that the
algebraic Bethe ansatz method can be applied as in \cite{Bax73,
Hou89, Tak92, Fel96, Hou03}.

Let us adopt the standard notation for convenience: \bea
&&\A(u)=\tilde{L}^1_1(u),~~\B_i(u)=\tilde{L}^1_i(u),~~i=2,
\cdots,n,\label{Con1}\\
&&\C_i(u)=\tilde{L}^i_1(u),~~~\D^{j}_{i}(u)=\tilde{L}^j_i(u),
~~i,j=2,\cdots,n.\label{Con2}\eea The transfer matrices $T(u)$
become \bea
T(u)=\A(u)+\sum_{i=2}^{n}\D^i_i(u).\label{transfer1}\eea Any
non-zero vector $|\O\rangle\in Fun(W)[\L]$ is of form
$|\O\rangle=g(\l)|\L\rangle$, for some meromorphic function $g\ne
0$. When $N=n\times l$, the weight $\L$ can be  written in the
form \bea \L=nl\L_1=l\sum_{k=1}^{n-1}(\e_1-\e_{k+1}).\eea Noting
(\ref{Exchange2}), the zero-weight vector space is spanned by the
vectors of the following form \bea
 \B_{i_{N_1}}(v_{N_1})\B_{i_{N_1-1}}(v_{N_1-1})\cdots\B_{i_1}(v_1)
|\O\rangle, \label{Zero} \eea where $N_1=(n-1)\times l$ and among
the indices $\lt\{i_k~|~k=1,\cdots,N_1\rt\}$, the number of
$i_k=j$, denoted
 by $\#(j)$, should be  \bea
 \#(j)=l, ~~~j=2, \cdots, n.\label{zero-weight}\eea
The above  states (\ref{Zero})  actually  belong to the
zero-weight space $W[0]$ \cite{Hou03}. Let us introduce the
following set of integers for convenience: \bea N_{i}=(n-i)\times
l,~~i=1,2,\cdots,n-1,\label{Integer}\eea and $\frac{n(n-1)}{2}l$
parameters
$\{\{v^{(i)}_k|~k=1,2,\cdots,N_{i+1}\},~i=0,1,\cdots,n-2\}$. The
parameters $\{\{v^{(i)}_k\}\}$ will be specified later by the
Bethe ansatz equations (\ref{BAE}). We will seek the common
eigenvectors of the {\it transfer matrices} $T(u)$ in the form
\bea |\l;\{v_k\}\rangle =\sum_{i_1,\cdots,i_{N_1}=2}^n
F^{i_1,i_2,\cdots,i_{N_1}}(\l;\lt\{v_k\rt\})
\B_{i_{N_1}}(v_{N_1})\B_{i_{N_1-1}}(v_{N_1-1})
\cdots\B_{i_1}(v_1)|\O\rangle,\no\\
~~\label{Eigenstate}\eea with the restriction condition
(\ref{zero-weight}). We adopt hereafter the convention: \bea
v_k=v^{(0)}_k,~k=1,2,\cdots, (n-1)\times l.\eea Let us introduce
$n$ parameters $\{\alpha^{(i)}|i=1,\cdots n\}$ to specify
quasi-vacua of each step of the nested Bethe ansatz \cite{Hou03},
and another set of parameters related to them: \bea
\bar{\alpha}^{(i)}=\frac{1}{(n-i-1)}\lt\{\alpha^{(i+1)}-\frac
{\sum_{k=i+1}^n\alpha^{(k)}}{n-i}\rt\}, ~~i=0,\cdots,
n-2.\label{Relation} \eea Choosing the function of $g(\l)$
 \bea
g(\l)=e^{\sqrt{-1}\pi \langle
\alpha^{(1)}\e_1,~\l\rangle}\prod_{j=2}^{n}\lt\{
\prod_{k=1}^{l}\frac{\s(\l_{j1}+kw)}{\s(w)}\rt\},\label{g-fun}\eea
then we have
\begin{Theorem} \cite{Hou03} With properly chosen
coefficients $F^{i_1,i_2,\cdots,i_{N_1}}(\l;\{v_k\})$, we obtain
eigenvectors of the transfer matrices \bea T(u)|\l ;\{v_k\}\rangle
=t(u;\{v_k\})|\l;\{v_k\} \rangle,\label{E-Pr}\eea with the
eigenvalue \bea t(u;\{v_k\})&=& e^{\sqrt{-1}\pi(1-n)\a w}
\lt\{\prod_{k=1}^{N_1}\frac{\s(v_k-u+w)}{\s(v_k-u)}\rt\}\no\\
&&~~~+e^{\sqrt{-1}\pi\a w}
\lt\{\prod_{k=1}^{N_1}\frac{\s(u-v_k+w)}{\s(u-v_k)}\rt\}
\lt\{\prod_{k=1}^{m}\frac{\s(u-p_k)}{\s(u-q_k)}\rt\}
t^{(1)}(u;\{v^{(1)}_k\}).\no\\
\label{t-function}\eea The functions $t^{(i)}(u;\{v^{(i)}_{k}\})$
are given recursively \bea &&t^{(i)}(u;\{v^{(i)}_k\})=
e^{\sqrt{-1}\pi(i+1-n)\a^{(i)} w}
\lt\{\prod_{k=1}^{N_{i+1}}\frac{\s(v^{(i)}_k-u+w)}{\s(v^{(i)}_k-u)}\rt\}\no\\
&&~~~~~~+e^{\sqrt{-1}\pi\a^{(i)} w}
\lt\{\prod_{k=1}^{N_{i+1}}\frac{\s(u-v^{(i)}_k+w)}{\s(u-v^{(i)}_k)}\rt\}
\lt\{\prod_{k=1}^{N_i}\frac{\s(u-p^{(i)}_k)}{\s(u-q^{(i)}_k)}\rt\}
t^{(i+1)}(u;\{v^{(i+1)}_k\}),\no\\
&&~~~~~~~~~~~~~~~~~~~~~~~~~~~~~~~~i=0,1,\cdots,n-2,\label{Recur1}\\
&&t^{(n-1)}(u)=1,~~ t^{(0)}(u;\{v^{(0)}_k\})
=t(u;\{v_k\}),\label{Recur2} \eea where $\a^{(i)}$, $i=0,1,\cdots,
n-2$ are given by (\ref{Relation}), $\a^{(0)}=\a$, $N_0=m$ and
\bea
&&p^{(0)}_k=p_k=z_k,~~q^{(0)}_k=q_k=z_k-l_kw,~~k=1,2,\cdots,m,
\\
&&p^{(i)}_k=v^{(i-1)}_k,~~q^{(i)}_k=v^{(i-1)}_k-w,~~i=1,2,\cdots,
n-2,~~~k=1,2,\cdots,N_{i}.\label{P-con}\eea The
$\{\{v^{(i)}_k\}\}$ satisfy the following Bethe ansatz equations
\bea e^{\sqrt{-1}\pi(i-n)\a^{(i)} w} \lt\{ \prod_{k=1,k\ne
s}^{N_{i+1}}\frac{\s(v^{(i)}_k-v^{(i)}_s+w)}{\s(v^{(i)}_k-v^{(i)}_s-w)}\rt\}=
\lt\{
\prod_{k=1}^{N_i}\frac{\s(v^{(i)}_s-p^{(i)}_k)}{\s(v^{(i)}_s-q^{(i)}_k)}\rt\}
t^{(i+1)}(v^{(i)}_s;\{v^{(i+1)}_k\}).\no\\
\label{BAE}\eea
\end{Theorem}

\vspace{0.6truecm}

\noindent In principle one can  construct explicit expression of
the coefficients of $F^{i_1,i_2,\cdots,i_{N_1}}(\l;\{v_k\})$ (for
details we refer the reader to \cite{Hou03}).

We conclude this section with some remarks on functional
dependence of the states $|\l;\{v_k\}\rangle$. By construction
(\ref{Con1}), the operators $\{\B_i\}$  are the functions of
$\{\l_i-\l_j\}$ because of the definition of the R-matrix
(\ref{Elements1})--(\ref{Elements2}), and the states can be
written in the following form \bea
|\l;\{v_k\}\rangle=\exp\{\sum_{i=1}^n\sqrt{-1}\pi\alpha^{(i)}\l_i\}
\overline{|\l;\{v_k\}\rangle}.\label{Properties1}\eea Here
$\overline{|\l;\{v_k\}\rangle}$ is a meromorphic function of
$\{\l_i\}$ and has the following properties \bea
&&\overline{|\l_1+c,\cdots,\l_n+c;\{v_k\}\rangle}=
\overline{|\l_1,\cdots,\l_n;\{v_k\}\rangle}, ~~{\rm for}~\forall
c\in
\Cb,\\
&&\overline{|\l_1,\cdots,\l_{i-1},\l_i+1,\l_{i+1},\cdots\l_n;\{v_k\}\rangle}=
(-1)^{l(n-1)}\overline{|\l_1,\cdots,\l_n;\{v_k\}\rangle},
~~i=1,\cdots,n. \no\\
\label{Properties2}\eea

\section{Ruijsenaars-Schneider models associated with $A_{n-1}$ root system }
\label{RS} \setcounter{equation}{0}
\subsection{The elliptic case}

Let us choose an $\E$-module $W$, a special one which corresponds
to the Young diagram
$\stackrel{nl}{\overbrace{\Box\!\Box\!\Box\cdots\Box}}$: \bea
W=V_{\L^{(nl)}}(0),\eea in which the evaluation point $z_1$ is set
to $0$. Then the zero-weight space of this module is
one-dimensional: $
|\l;\{v_k\}\rangle=\tilde{\Phi}_{RS}(\l;\{\alpha^{(i)}\})e_0,~~e_0\in
W[0]$ and it does not depend on $\l_i$. The associated {\it
transfer matrices} can be written as \cite{Fel961} \bea
T(u)=\frac{\s(u+lw)}{\s(u+nlw)}M.\label{Ham}\eea The operator $M$
is independent of $u$ and is given by \bea M=\sum_{i=1}^n\lt\{
\prod_{j\ne i}\frac{\s(\l_{ij}+lw)}{\s(\l_{ij})}\G_i\rt\}.\eea
Here $\lt\{\G_i\rt\}$ are difference operators:
$\G_i~f(\l)=f(\l-w\hat{i})$. Noting
(\ref{Properties1})--(\ref{Properties2}) and (\ref{Relation1}), we
find \bea M\tilde{\Phi}_{RS}= \tilde{H}_{RS}\tilde{\Phi}_{RS}=
\e_{RS} \tilde{\Phi}_{RS}.\label{E-V}\eea The difference operator
$\tilde{H}_{RS}$ is given by  \bea
\tilde{H}_{RS}=\sum_{i=1}^n\lt\{ \prod_{j\ne
i}\frac{\s(\l_{ij}+lw)}{\s(\l_{ij})}e^{-w\frac{\partial}{\partial
\l_i}}\rt\}.\eea

In order to apply Theorem 2 to RS model, hereafter we restrict the
parameters $\tau$ and $w$ as follows: \bea
\tau=\sqrt{-1}\k,~\k\in\R,~\k>0,~~w=\w,\label{Paramters} \eea
where $g$ is a real number. This is necessary for the reality of
the Hamiltonian. Because the parameters $\{\l_i\}$ will play the
role of the canonical coordinates, we further restrict
$\l_i\in\R$. In terms of the specified parameters,
$\tilde{H}_{RS}$ becomes \bea \tilde{H}_{RS}=\sum_{i=1}^n\lt\{
\prod_{j\ne i}\frac{\s(\l_{ij}+\w
l)}{\s(\l_{ij})}e^{-\w\frac{\partial}{\partial
\l_i}}\rt\}.\label{Rui-elliptic}\eea The resulting difference
operator $\tilde{H}_{RS}$ will be  the Hamiltonian of elliptic
$A_{n-1}$ type RS model \cite{Rui87} with the special {\it
coupling constant\/} $\g=\w l$, up to conjugation by a function
\cite{Has97, Hou00}.  Suppose $\tilde{\Phi}_{RS}$ and $\e_{RS}$
are an eigenfunction and the corresponding eigenvalue of
$\tilde{H}_{RS}$ \bea
\tilde{H}_{RS}\tilde{\Phi}_{RS}=\e_{RS}\tilde{\Phi}_{RS}.\eea Let
us introduce another function $\Phi_{RS}$ \bea
\Phi_{RS}=e^{-\Psi_{RS}}\tilde{\Phi}_{RS},~~\Psi_{RS}=\frac{1}{2}
\ln\prod_{i\ne j}\lt\{\prod_{k=1}^l\frac{\s(\l_{ij}-\w k)}{\s(\w
)}\rt\},\label{Transformation}\eea associated to
$\tilde{\Phi}_{RS}$. Then $\Phi_{RS}$ is an eigenfunction of the
similarity transformed Hamiltonian $H_{RS}$ with the same
eigenvalue $\e_{RS}$ \bea &&H_{RS}\Phi_{RS}=\e_{RS}\Phi_{RS},~~~
H_{RS}=e^{-\Psi_{RS}}\tilde{H}_{RS}e^{\Psi_{RS}},\\
&&H_{RS}=\sum_{i=1}^n\lt\{\prod_{j\ne i}\frac{\s(\l_{ji}-\w
l)}{\s(\l_{ji})}\rt\}^{\frac{1}{2}}~e^{-\w\frac{\partial}{\partial
\l_i}}~\lt\{\prod_{j\ne i}\frac{\s(\l_{ij}-\w
l)}{\s(\l_{ij})}\rt\}^{\frac{1}{2}}. \label{H}\eea One finds that
$H_{RS}$ is the  Hamiltonian of elliptic $A_{n-1}$ type RS model
\cite{Rui87} with the special {\it coupling constant\/} $\g=\w l$.

Now, we consider the spectrum of $\tilde{H}_{RS}$. Theorem 2
enables us to obtain the spectrum of the Hamiltonian of the
elliptic $A_{n-1}$ Ruijsenaars-Schneider model as well as the
eigenfunctions, in terms of  the associated {transfer matrices}
(\ref{Ham}). Since we have already taken the special $\E$-module
$W=V_{\L^{(n\times l)}}(0)$, thanks to Theorem 2, the eigenvalues
are given by (\ref{E-Pr}) and (\ref{t-function}) but with special
values of the parameters \bea
m=N_0=1,~~p^{(0)}_1=z_1=0,~~q^{(0)}_1=-\w nl.\label{One}\eea Since
$M$ is independent of $u$, we can evaluate the eigenvalues of
$T(u)$ at $u=z_1=0$. Then the expression of the eigenvalue
$t(u;\{v_k\})$ simplifies drastically, for the second term in the
right hand side of (\ref{t-function}) (the one depending on the
eigenvalue of the reduced transfer matrices
$t^{(1)}(u;\{v_k^{(1)}\})$) vanishes because $u-p^{(0)}_1=0$.

Note that $\tilde{H}_{RS}$ ($H_{RS}$) has periodic coefficients
with $\l_i\rightarrow \l_i+1$, and therefore preserves the space
of Bloch functions such that \bea
\psi(\l_1,\cdots,\l_{i-1},\l_i+1,\l_{i+1},\cdots,\l_n)=\pm
(-1)^{l(n-1)}\psi(\l_1,\cdots,\l_n).\label{Bloch}\eea The (quasi)
periodicity requires integer $\alpha^{(i)}$, $\alpha^{(i)}\in\Zb$.
Noting the relation (\ref{Relation}), in order to get one-to-one
correspondence between $\{\alpha^{(i)}\}$ and $\{\a^{(i)}\}$, we
need further choose \bea
\alpha^{(i)}\in\Zb^{+},~~i=1,\cdots,n-1,~~ {\rm and}~
\alpha^{(n)}=-\sum_{k=1}^{n-1}\alpha^{(k)}.\label{Relation1}\eea

Finally, we obtain the eigenvalues $ \e_{RS}(\lt\{v_k\rt\})$ of
$\tilde{H}_{RS}$ (\ref{E-V}): \bea e^{\pi(n-1)\a g}\frac{\s(\w
nl)}{\s(\w l)}\lt\{\prod_{k=1}^{(n-1)\times l} \frac {\s(v_k+\w)}
{\s(v_k)} \rt\},\eea where $\lt\{\lt\{v^{(i)}_k\rt\}\rt\}$ satisfy
the Bethe ansatz equations \bea &&e^{\pi(n-i)\a^{(i)}g} \lt\{
\prod_{k=1,k\ne
s}^{N_{i+1}}\frac{\s(v^{(i)}_k-v^{(i)}_s+\w)}{\s(v^{(i)}_k-v^{(i)}_s-\w)}\rt\}=
\lt\{
\prod_{k=1}^{N_i}\frac{\s(v^{(i)}_s-p^{(i)}_k)}{\s(v^{(i)}_s-q^{(i)}_k)}\rt\}
t^{(i+1)}(v^{(i)}_s;\lt\{v^{(i+1)}_k\rt\}),\no\\
&&~~~~~~~~i=0,1,\cdots, n-2.\label{BAE2}\eea The functions
$t^{(i)}(u)$ appearing in (\ref{BAE2})  are given by the same
recurrence relations as (\ref{Recur1})--(\ref{Recur2}), but with
the special $N_0=1$, $p^{(0)}_1=0$ and  $q^{(0)}_1=-\w nl$ and
replacing  $w$ by $\w$. Substituting the expression of the
function $t^{(i+1)}(u)$ (\ref{Recur1}) into the Bethe ansatz
equations (\ref{BAE2}), noting the conditions (\ref{One}) and
(\ref{P-con}), we have
\begin{Proposition}
The eigenvalues of the Hamiltonian (\ref{H}) of the elliptic RS
model associated with  $A_{n-1}$ root system with the discrete
coupling constant $\g=\w l$ ($l$ being an integer) are \bea
\e_{RS}=e^{\pi(n-1)\a g}\frac{\s(\w nl)}{\s(\w
l)}\lt\{\prod_{k=1}^{(n-1)\times l} \frac {\s(v_k +\w )} {\s(v_k)}
\rt\}.\label{Eig-R}\eea The $\frac{n(n-1)}{2}l$ parameters
$\{\{v^{(i)}_k\}\}$ satisfy the Bethe ansatz equations \bea
&&e^{\pi n\a g} \prod_{k=1,k\ne
s}^{N_{1}}\frac{\s(v_k-v_s+\w)}{\s(v_k-v_s-\w)}=
e^{\pi(n-2)\a^{(1)}g}~\frac{\s(v_s)}{\s(v_s+\w nl)}\no\\
&&~~~~~~~~~~~~~~~~~~~~~~~~~~~~~~~~\times
\prod_{k=1}^{N_2}\frac{\s(v^{(1)}_k-v_s+\w)}{\s(v^{(1)}_k-v_s)},
\label{BA-R1}\\
&&e^{\pi(n-i)\a^{(i)}g} \prod_{k=1,k\ne
s}^{N_{i+1}}\frac{\s(v^{(i)}_k-v^{(i)}_s+\w)}{\s(v^{(i)}_k-v^{(i)}_s-\w)}=
e^{\pi(n-i-2)\a^{(i+1)}g}~
\prod_{k=1}^{N_i}\frac{\s(v^{(i)}_s-v^{(i-1)}_k)}{\s(v^{(i)}_s-v^{(i-1)}_k+\w)}\no\\
&&~~~~~~~~~~~~~~~~~~~~~~~~~~~~~~~~\times
\prod_{k=1}^{N_{i+2}}\frac{\s(v^{(i+1)}_k-v^{(i)}_s+\w)}{\s(v^{(i+1)}_k-v^{(i)}_s)}
,~~i=1,\cdots, n-3,\\
&&e^{2\pi\a^{(n-2)}g}  \prod_{k=1,k\ne
s}^{N_{n-1}}\frac{\s(v^{(n-2)}_k-v^{(n-2)}_s+\w)}{\s(v^{(n-2)}_k-v^{(n-2)}_s-\w)}=
\prod_{k=1}^{N_{n-2}}\frac{\s(v^{(n-2)}_s-v^{(n-3)}_k)}
{\s(v^{(n-2)}_s-v^{(n-3)}_k+\w)}.\label{BA-R2} \eea The parameters
$\a^{(i)}$, $i=0,1,\cdots, n-2$ and $\a^{(0)}=\a$ are given by the
relation (\ref{Relation}) from $n-1$ arbitrary non-negative
integers $\{\alpha^{(i)}\in\Zb^{+}|i=1,\cdots,n-1\}$.
\end{Proposition}
\vspace{0.6truecm}

Our formula of eigenvalues is the elliptic generalization of the
{\it third formulas} in the sense of Felder et al \cite{Fel95}.
Taking complex conjugation of the Bethe ansatz equations
(\ref{BA-R1})-(\ref{BA-R2}), noting the property (\ref{real}), we
find that the solutions $\{\{v^{(i)}_k\}\}$ are all pure imaginary
numbers. This ensures that the eigenvalues $\e_{RS}$ are real. By
construction from the nested Bethe ansatz method and the relation
(\ref{Transformation}), we know that the corresponding
eigenfunction is a meromorphic function of $\{\l_i\}$ and
satisfies the following properties \bea
\Phi_{RS}(\l_1,\cdots,\l_{i-1},\l_i+1,\l_{i+1},\cdots,\l_n;
\{\alpha^{(i)}\}) =(-1)^{\alpha^{(i)}}\Phi_{RS}
(\l;\{\alpha^{(i)}\}).\label{Qasi-P}\eea

\subsection{The trigonometric case}
Here we consider the trigonometric RS model associated with
$A_{n-1}$ root system. The corresponding Hamiltonian with the
discrete {\it coupling constant} $\g=\w l$  is given \bea
H_{RS}=\sum_{i=1}^n\lt\{ \prod_{j\ne i}\frac{\sin\pi(\l_{ji}-\w
l)}{\sin\pi(\l_{ji})}\rt\}^{\frac{1}{2}}e^{-\w\frac{\partial}{\partial
\l_i}}\lt\{ \prod_{j\ne i}\frac{\sin\pi(\l_{ij}-\w
l)}{\sin\pi(\l_{ij})}\rt\}^{\frac{1}{2}}.\label{Ham-tri}\eea

Taking the trigonometric limit $\k\rightarrow +\infty$
($\tau\rightarrow +\sqrt{-1}\infty)$, one finds that \bea
\frac{\s(x)}{\s(y)}~~\longrightarrow~\frac{\sin\pi x}{\sin\pi
y},~{\rm when }~\k\rightarrow +\infty, \eea from the product
expression (\ref{A1}) of the $\s$-function. The trigonometric
Hamiltonian (\ref{Ham-tri}) can be obtained from the elliptic one
(\ref{H}) by taking limit $\k\rightarrow +\infty$. Consequently,
we can find the spectrum of the Hamiltonian of the trigonometric
RS model associated with $A_{n-1}$ type root system by taking the
trigonometric limit of the elliptic one. Noting that the solutions
to the Bethe ansatz equations (\ref{BA-R1})--(\ref{BA-R2})
$\{\{v^{(i)}_k\}\}$ are all pure imaginary numbers, let us
introduce $\frac{n(n-1)}{2}l$ real parameters
$\{\{\bar{v}^{(i)}_k\}\}$ associated to $\frac{n(n-1)}{2}l$ pure
imaginary parameters $\{\{v^{(i)}_k\}\}$ as follows: \bea
v^{(i)}_{k}=\sqrt{-1}\bar{v}^{(i)}_k, ~~ \bar{v}^{(i)}_k\in
\mathbb{R}.\label{Par}\eea Finally, we have

\begin{Proposition}
The eigenvalues of the Hamiltonian (\ref{Ham-tri}) of the
trigonometric  RS model associated with $A_{n-1}$ root system with
the discrete coupling constant $\g=\w l$  are \bea
\e_{RS}=e^{\pi(n-1)\a g}\frac{\sinh\pi(nlg)}{\sinh\pi(lg)}
\lt\{\prod_{k=1}^{(n-1)\times l} \frac
{\sinh\pi(\bar{v}^{(0)}_k+g)} {\sinh\pi(\bar{v}^{(0)}_k)}
\rt\}.\eea The $\frac{n(n-1)}{2}l$ real parameters
$\{\{\bar{v}^{(i)}_k\}\}$ satisfy the Bethe ansatz equations \bea
&&e^{\pi n\a g} \lt\{ \prod_{k=1,k\ne
s}^{N_{1}}\frac{\sinh\pi(\bar{v}^{(0)}_k-\bar{v}^{(0)}_s+g)}
{\sin\pi(\bar{v}^{(0)}_k-\bar{v}^{(0)}_s-g)}\rt\}=
e^{\pi(n-2)\a^{(1)}g}~\frac{\sinh\pi(\bar{v}^{(0)}_s)}
{\sinh\pi(\bar{v}^{(0)}_s+nlg)}\no\\
&&~~~~~~~~~~~~~~~~~~~\times \lt\{
\prod_{k=1}^{N_2}\frac{\sinh\pi(\bar{v}^{(1)}_k-\bar{v}^{(0)}_s+g)}
{\sinh\pi(\bar{v}^{(1)}_k-\bar{v}^{(0)}_s)}\rt\},\\
&&e^{\pi(n-i)\a^{(i)}g} \lt\{ \prod_{k=1,k\ne
s}^{N_{i+1}}\frac{\sinh\pi(\bar{v}^{(i)}_k-\bar{v}^{(i)}_s+g)}
{\sinh\pi(\bar{v}^{(i)}_k-\bar{v}^{(i)}_s-g)}\rt\}=
e^{\pi(n-i-2)\a^{(i+1)}g}\no\\
&&~~~~~~~~~~~~~~~~~~~\times \lt\{
\prod_{k=1}^{N_i}\frac{\sinh\pi(\bar{v}^{(i)}_s-\bar{v}^{(i-1)}_k)}
{\sinh\pi(\bar{v}^{(i)}_s-\bar{v}^{(i-1)}_k+g)}\rt\}\lt\{
\prod_{k=1}^{N_{i+2}}\frac{\sinh\pi(\bar{v}^{(i+1)}_k-\bar{v}^{(i)}_s+g)}
{\sinh\pi(\bar{v}^{(i+1)}_k-\bar{v}^{(i)}_s)}\rt\},\no\\
&&~~~~~~~~~~~~~~~~~~~~~~~~i=1,\cdots, n-3,\\
&&e^{2\pi\a^{(n-2)} g} \lt\{ \prod_{k=1,k\ne
s}^{N_{n-1}}\frac{\sinh\pi(\bar{v}^{(n-2)}_k-\bar{v}^{(n-2)}_s+g)}
{\sinh\pi(\bar{v}^{(n-2)}_k-\bar{v}^{(n-2)}_s-g)}\rt\}=
\lt\{\prod_{k=1}^{N_{n-2}}\frac{\sinh\pi(\bar{v}^{(n-2)}_s-\bar{v}^{(n-3)}_k)}
{\sinh\pi(\bar{v}^{(n-2)}_s-\bar{v}^{(n-3)}_k+g)}\rt\}.\no\\
 \eea Here the parameters $\a^{(i)}$, $i=0,1,\cdots, n-2$ and $\a=\a^{0}$
are given by the relation (\ref{Relation}) from $n-1$ arbitrary
non-negative integers $\{\alpha^{(i)}\in\Zb^{+}|i=1,\cdots,n-1\}$.
\end{Proposition}
\vspace{0.6truecm}

Our formula of eigenvalues is the {\it trigonometric}
generalization of the {\it third formulas} in the sense of Felder
et al \cite{Fel95}. Similarly to the elliptic case, we know that
the corresponding eigenfunction has the same quasi-periodic
properties (\ref{Qasi-P}).

\subsection{The rational case}
The Hamiltonian of the rational RS model associated with $A_{n-1}$
root system reads as \bea H_{RS}=\sum_{i=1}^n\lt\{ \prod_{j\ne
i}(1-\frac{\w
l}{\l_{ji}})\rt\}^{\frac{1}{2}}e^{-\w\frac{\partial}{\partial
\l_i}}\lt\{ \prod_{j\ne i}(1-\frac{\w
l}{\l_{ij}})\rt\}^{\frac{1}{2}}.\label{R-Ham}\eea If one rescales
\bea &&\l_i~\longrightarrow~\d\,\l_i,
~~\a^{(i)}~\longrightarrow~\frac{1}{\pi\d}\,\a^{(i)},
~~g~\longrightarrow~\d\,g,\label{Rscale1}\\
&&\bar{v}^{(i)}_k~\longrightarrow~\d\,\bar{v}^{(i)}_k,\label{Rscale2}\eea
and takes the limit: $\d~\longrightarrow~0$ (we call it the
rational limit), the Hamiltonian (\ref{R-Ham}) of the rational RS
model can be obtained from the trigonometric one (\ref{Ham-tri}).
Therefore we can find the spectrum of the Hamiltonian of the
rational RS model associated with $A_{n-1}$ root system by taking
the rational limit of Proposition 2.  Finally we have

\begin{Proposition}
The eigenvalues of the Hamiltonian (\ref{R-Ham}) of the rational
RS model associated with $A_{n-1}$ root system with the discrete
coupling constant $\g=\w l$  are \bea \e_{RS}=n e^{(n-1)\a
g}\lt\{\prod_{k=1}^{(n-1)\times l} \frac {\bar{v}^{(0)}_k+g}
{\bar{v}^{(0)}_k} \rt\}.\eea The $\frac{n(n-1)}{2}l$ real
parameters $\{\{\bar{v}^{(i)}_k\}\}$ satisfy the Bethe ansatz
equations \bea &&e^{n\a g} \lt\{ \prod_{k=1,k\ne
s}^{N_{1}}\frac{(\bar{v}^{(0)}_k-\bar{v}^{(0)}_s+g)}
{(\bar{v}^{(0)}_k-\bar{v}^{(0)}_s-g)}\rt\}=
e^{(n-2)\a^{(1)}g}~\frac{\bar{v}^{(0)}_s}{\bar{v}^{(0)}_s+nlg}
\lt\{ \prod_{k=1}^{N_2}\frac{(\bar{v}^{(1)}_k-\bar{v}_s+g)}
{(\bar{v}^{(1)}_k-\bar{v}_s)}\rt\},\no\\
\\
&&e^{(n-i)\a^{(i)} g} \lt\{ \prod_{k=1,k\ne
s}^{N_{i+1}}\frac{(\bar{v}^{(i)}_k-\bar{v}^{(i)}_s+g)}
{(\bar{v}^{(i)}_k-\bar{v}^{(i)}_s-g)}\rt\}=
e^{(n-i-2)\a^{(i+1)}g}~ \lt\{
\prod_{k=1}^{N_i}\frac{(\bar{v}^{(i)}_s-\bar{v}^{(i-1)}_k)}
{(\bar{v}^{(i)}_s-\bar{v}^{(i-1)}_k+g)}\rt\}\no\\
&&~~~~~~~~~~~~~~~~~~~~~~~~~~~~~~\times \lt\{
\prod_{k=1}^{N_{i+2}}\frac{(\bar{v}^{(i+1)}_k-\bar{v}^{(i)}_s+g)}
{(\bar{v}^{(i+1)}_k-\bar{v}^{(i)}_s)}\rt\}
,~~i=1,\cdots, n-3,\\
&&e^{2\a^{(n-2)}g} \lt\{ \prod_{k=1,k\ne
s}^{N_{n-1}}\frac{(\bar{v}^{(n-2)}_k-\bar{v}^{(n-2)}_s+g)}
{(\bar{v}^{(n-2)}_k-\bar{v}^{(n-2)}_s-g)}\rt\}=
\lt\{\prod_{k=1}^{N_{n-2}}\frac{(\bar{v}^{(n-2)}_s-\bar{v}^{(n-3)}_k)}
{(\bar{v}^{(n-2)}_s-\bar{v}^{(n-3)}_k+g)}\rt\}. \eea Here the
parameters $\a^{(i)}$, $i=0,1,\cdots, n-2$ and $\a^{(0)}=\a$ are
given by the relation (\ref{Relation}) from $n-1$ arbitrary
non-negative real numbers
$\{\alpha^{(i)}\in\R^{+}|i=1,\cdots,n-1\}$.
\end{Proposition}
\vspace{0.6truecm}

Our formula of eigenvalues is the rational generalization of the
{\it third formulas} in the sense of Felder et al \cite{Fel95}.
The rescalings (\ref{Rscale1}) and (\ref{Rscale2}) and the
rational limit lead to that the coefficients of the Hamiltonian
(\ref{R-Ham}) are no longer periodic (cf. the elliptic and
trigonometric case). The (quasi) periodic properties
(\ref{Qasi-P}) of the eigenfunction now are replaced by the
following asymptotic properties: \bea \Phi_{RS}\propto
e^{\sqrt{-1}\alpha^{i}\l_i},~~\l_i\rightarrow
\infty.\label{asym}\eea Namely, the corresponding eigenfunctions
are bounded when $\l_i\longrightarrow \infty$ on the real axis.

\section{Calogero-Moser systems associated with $A_{n-1}$ root system}
\label{CM} \setcounter{equation}{0}

In this section, we will study all types of CM models associated
with $A_{n-1}$ root system by taking  ``non-relativistic" limit
\cite{Bra97}: $g\rightarrow 0$ of the corresponding RS models
which have already been studied in the Sect. 3.
\subsection{Elliptic potential}
Taking the {\it non-relativistic limit} of the Hamiltonian
(\ref{Rui-elliptic}) and noting the asymptotic properties of
$\s$-function (\ref{P1}), we obtain  \bea
\tilde{H}_{RS}=n+\frac{g^2}{2}\tilde{H}_{CM}+O(g^3),~~~g\rightarrow
0. \label{PHam}\eea The resulting differential operator
$\tilde{H}_{CM}$ is given by \bea \tilde{H}_{CM}=- \sum_{i=1}^n
\frac{\partial^2}{(\partial\l_i)^2}+2l\sum_{i\ne j}^n\z(\l_{ij})
\frac{\partial}{\partial\l_i}-l^2\sum_{i\ne
j}\frac{\s''(\l_{ij})}{\s(\l_{ij})}-l^2\sum_{i\ne j\ne k}
\z(\l_{ij})\z(\l_{ik}),\eea where
$\s''(t)=\frac{\partial^2}{(\partial u)^2}\s(u)|_{u=t}$ and the
function $\z$ is defined in (\ref{Function}). We can further
transform  $\tilde{H}_{CM}$ to a more familiar form.  Let us
suppose $\tilde{\Phi}$ and $\e_{CM}$ are an eigenfunction and the
corresponding eigenvalue of $\tilde{H}_{CM}$, namely,  \bea
\tilde{H}_{CM}\tilde{\Phi}=\e_{CM}\tilde{\Phi}.\eea At the same
time, we introduce another function $\Phi$ \bea
\Phi=e^{-\Psi}\tilde{\Phi},~~\Psi=\ln\prod_{i<
j}\lt(\s(\l_{ij})\rt)^l,\eea associated to $\tilde{\Phi}$. Then
$\Phi$ is an eigenfunction of the Hamiltonian  $H_{CM}$ with the
same eigenvalue $\e_{CM}$ \bea H_{CM}\Phi=\e_{CM}\Phi,~~~
H_{CM}=e^{-\Psi}\tilde{H}_{CM}e^{\Psi}= -\sum_{i=1}^n
\frac{\partial^2}{(\partial\l_i)^2}+l(l+1)\sum_{i\ne
j}\p(\l_{ij}),\label{Ham-CM}\eea where the function $\p$ is
defined in (\ref{Function}). One finds that $H_{CM}$ is exactly
the Hamiltonian of elliptic CM model associated with  $A_{n-1}$
root system \cite{Cal75,Mos75} with the {\it coupling constant}
$l+1$ \footnote{Traditionally, the {\it coupling constant} of the
Hamiltonian: $ H_{CM}=-\sum_{i=1}^n
\frac{\partial^2}{(\partial\l_i)^2}+\g(\g-1)\sum_{i\ne
j}\p(\l_{ij}) $ of CM model is set to $\g$.}.

Now we study the asymptotic properties of the eigenvalues of
$\tilde{H}_{RS}$ (\ref{Eig-R}) and the associated Bethe ansatz
equations (\ref{BA-R1})--(\ref{BA-R2}). Let the solution to the
Bethe ansatz equations (\ref{BA-R1})-(\ref{BA-R2}) have the
following form\bea v^{(i)}_k=x^{(i)}_k+\w
y^{(i)}_k-g^2z^{(i)}_k+O(g^3),~~g\rightarrow 0.\eea Noting the
asymptotic properties of $\s$-function (\ref{P1}), the equation
(\ref{Eig-R}) becomes \bea \e_{RS}&=&n+\w
n\lt((1-n)\sqrt{-1}\pi\a+\sum_{k=1}^{N_1}\z(x^{(0)}_k)\rt)\no\\
&&~~~~+\frac{g^2}{2}\lt\{n\sum_{k=1}^{N_1}(2y^{(0)}_k+1)\p(x^{(0)}_k)
-\frac{(n+1)n(n-1)l^2}{3}~\frac{\s'''(0)}{\s'(0)}\rt.\no\\
&&~~~~-\lt.n\lt((1-n)\sqrt{-1}\pi\a+\sum_{k=1}^{N_1}\z(x^{(0)}_k)\rt)^2
\rt\} +O(g^3).\label{Eig-CM}\eea

The Bethe ansatz equations (\ref{BA-R1})--(\ref{BA-R2}) at the
first order of $g$ become \bea &&2\sum_{k=1,k\ne
s}^{N_1}\z(x^{(0)}_k-x^{(0)}_s)-n\sqrt{-1}\pi\a=
(2-n)\sqrt{-1}\pi\a^{(1)}-nl\z(x^{(0)}_s)\no\\
&&~~~~~~~~~~~~~~~~~~~~~~+\sum_{k=1}^{N_2}\z(x^{(1)}_k-x^{(0)}_s),
\label{BA-C1}\\
&&2\sum_{k=1,k\ne
s}^{N_{i+1}}\z(x^{(i)}_k-x^{(i)}_s)+(i-n)\sqrt{-1}\pi\a^{(i)}
=(i+2-n)\sqrt{-1}\pi\a^{(i+1)}\no\\
&&~~~~~~~~~~~~~~~~~~~~~~-
\sum_{k=1}^{N_i}\z(x_s^{(i)}-x^{(i-1)}_k)+\sum_{k=1}^{N_{i+2}}
\z(x^{(i+1)}_k-x^{(i)}_s),~~i=1,\cdots, n-3,\\
&&2\sum_{k=1,k\ne
s}^{N_{n-1}}\z(x^{(n-2)}_k-x^{(n-2)}_s)-2\sqrt{-1}\pi\a^{(n-2)}
=-\sum_{k=1}^{N_{n-2}}\z(x^{(n-2)}_s-x^{(n-3)}_k).
\label{BA-C2}\eea Sum up with $s$ for each equation of
(\ref{BA-C1})--(\ref{BA-C2}). Then taking the summation of  all
the equations and noting the parity property of $\z$-function
(\ref{Function}), we find \bea
\sum_{k=1}^{N_1}\z(x^{(0)}_k)+(1-n)\sqrt{-1}\pi\a=0.\label{R-1}\eea
This means that the first order of $g$ term of $\e_{RS}$ in
(\ref{Eig-CM}) is vanishing which is in conformity with
(\ref{PHam}).

The Bethe ansatz equations (\ref{BA-R1})--(\ref{BA-R2}) at the
second order of $g$ are  \bea &&4\sum_{k=1,k\neq
s}^{N_1}(y^{(0)}_s-y^{(0)}_k)\p(x^{(0)}_k-x^{(0)}_s)=nl(2y^{(0)}_s+nl)
\p(x^{(0)}_s)\no\\
&&~~~~~~~~~~~~~~~~-\sum_{k=1}^{N_2}(2y^{(1)}_k
-2y^{(0)}_s+1)\p(x^{(1)}_k-x^{(0)}_s),\label{S-1}\\
&&4\sum_{k=1,k\neq
s}^{N_{i+1}}(y^{(i)}_s-y^{(i)}_k)\p(x^{(i)}_k-x^{(i)}_s)=
\sum_{k=1}^{N_i}(2y^{(i)}_s
-2y^{(i-1)}_k+1)\p(x^{(i)}_s-x^{(i-1)}_k)\no\\
&&~~~~~~~~~~~~~~~~-\sum_{k=1}^{N_{i+2}}(2y^{(i+1)}_k
-2y^{(i)}_s+1)\p(x^{(i+1)}_k-x^{(i)}_s),~~i=1,\cdots,n-3,\\
&&4\sum_{k=1,k\neq
s}^{N_{n-1}}(y^{(n-2)}_s-y^{(n-2)}_k)\p(x^{(n-2)}_k-x^{(n-2)}_s)\no\\
&&~~~~~~~~~~~~~~~~= \sum_{k=1}^{N_{n-2}}(2y^{(n-2)}_s
-2y^{(n-3)}_k+1)\p(x^{(n-2)}_s-x^{(n-3)}_k).\label{S-2}\eea Sum up
with $s$ for each equation of (\ref{S-1})--(\ref{S-2}). Then
taking the summation of  all the equations and noting the parity
property of $\p$-function (\ref{Function}), we find \bea
\sum_{s=1}^{N_1}(2y^{(0)}_s+nl)\p(x^{(0)}_s)=0.\label{R-2}\eea

Substituting the equations (\ref{R-1}) and (\ref{R-2}) into
(\ref{Eig-CM}), we finally have

\begin{Proposition}
The eigenvalues of the Hamiltonian (\ref{Ham-CM}) of the elliptic
CM model associated with $A_{n-1}$ root system with the discrete
coupling constants $\g=l+1$  are \bea
\e_{CM}=(1-nl)n\sum_{k=1}^{N_1}\p(x^{(0)}_k)-\frac{(n+1)n(n-1)}{3}l^2
~\frac{\s'''(0)}{\s'(0)}.\eea The $\frac{n(n-1)}{2}l$ parameters
$\{\{x^{(i)}_k\}\}$ satisfy the Bethe ansatz equations \bea
&&2\sum_{k=1,k\ne s}^{N_1}\z(x^{(0)}_k-x^{(0)}_s)-n\sqrt{-1}\pi\a=
(2-n)\sqrt{-1}\pi\a^{(1)}-nl\z(x^{(0)}_s)\no\\
&&~~~~~~~~~~~~~~~~~~+\sum_{k=1}^{N_2}\z(x^{(1)}_k-x^{(0)}_s),
\label{BA-C3}\\
&&2\sum_{k=1,k\ne s}^{N_{i+1}}\z(x^{(i)}_k-x^{(i)}_s)
+(i-n)\sqrt{-1}\pi\a^{(i)}=(i+2-n)\sqrt{-1}\pi\a^{(i+1)}\no\\
&&~~~~~~~~~~~~~~~~~~-\sum_{k=1}^{N_i}\z(x_s^{(i)}-x^{(i-1)}_k)+\sum_{k=1}^{N_{i+2}}
\z(x^{(i+1)}_k-x^{(i)}_s),~~i=1,\cdots, n-3,\\
&&2\sum_{k=1,k\ne
s}^{N_{n-1}}\z(x^{(n-2)}_k-x^{(n-2)}_s)-2\sqrt{-1}\pi\a^{(n-2)}
=-\sum_{k=1}^{N_{n-2}}\z(x^{(n-2)}_s-x^{(n-3)}_k).
\label{BA-C4}\eea The parameters $\a^{(i)},~i=0,1,\cdots, n-2$ and
$\a^{(0)}=\a$ are given by the relation (\ref{Relation}) from
$n-1$ arbitrary non-negative integers
$\{\alpha^{(i)}\in\Zb^{+}|i=1,\cdots,n-1\}$.
\end{Proposition}

\vspace{0.6truecm}

Our result agrees with  the {\it third formulas} (or Bethe ansatz
type) of the eigenvalues of the elliptic CM model associated with
$A_{n-1}$ root system \cite{Fel95}. Taking  complex conjugation of
the Bethe ansatz equations (\ref{BA-C3})--(\ref{BA-C4}), noting
the property (\ref{P2}) and (\ref{real}), we find that the
solutions $\{\{x^{(i)}_k\}\}$ to the equations are all pure
imaginary numbers. This ensures that the eigenvalues $\e_{CM}$ are
real and positive up to {\it the ground state energy\/}
$\e_0=-\frac{(n+1)n(n-1)}{3}l^2 ~\frac{\s'''(0)}{\s'(0)}$ (the
positivity from the expression (\ref{Post}) of $\p$-function when
the argument is taken on imaginary axis).

\subsection{Trigonometric potential}
Here we consider  trigonometric CM  models associated with
$A_{n-1}$ root system. The corresponding Hamiltonian with the
discrete {\it coupling constant} $\g=l+1$ is given
 \bea H_{CM}=-\sum_{i=1}^n
\frac{\partial^2}{(\partial\l_i)^2}+l(l+1)\sum_{i\ne
j}\frac{\pi^2}{\sin^2(\pi\l_{ij})}. \label{Ham-Tr}\eea Taking the
trigonometric limit $\k\longrightarrow +\infty$, one finds that
\bea
&&\z(u)~\longrightarrow~\pi\cot\pi u,\label{Z-fuc}\\
&&\p(u)~\longrightarrow~\frac{\pi^2}{\sin^2(\pi
u)},\label{P-func}\eea from expansions (\ref{A2}) and (\ref{A3}).
Then the Hamiltonian (\ref{Ham-Tr}) can be obtained from the
elliptic type (\ref{Ham-CM}) by taking the trigonometric limit.
Moreover, since the solutions $\{\{x^{(i)}_k\}\}$ to the Bethe
ansatz equations (\ref{BA-C3})--(\ref{BA-C4}) are all pure
imaginary numbers, we can introduce $\frac{n(n-1)}{2}l$ real
parameters $\{\{\bar{x}^{(i)}_k\}\}$ associated with
$\{\{x^{(i)}_k\}\}$: \bea x^{(i)}_k=\sqrt{-1}\bar{x}^{(i)}_k.\eea
Finally, we can find the spectrum of the Hamiltonian of
trigonometric CM model associated with $A_{n-1}$ root system from
Proposition 4:
\begin{Proposition}
The eigenvalues of the Hamiltonian (\ref{Ham-Tr}) of the
trigonometric CM model associated with  $A_{n-1}$ root system with
the discrete coupling constant $\g=l+1$  are \bea
\e_{CM}=(nl-1)n\sum_{k=1}^{N_1}\frac{\pi^2}{\sinh^2(\pi\bar{x}^{(0)}_k)}
+\frac{(n+1)n(n-1)}{3}l^2\pi^2.\eea The  $\frac{n(n-1)}{2}l$ real
parameters $\{\{\bar{x}^{(i)}_k\}\}$ satisfy the Bethe ansatz
equations \bea &&2\sum_{k=1,k\ne
s}^{N_1}\coth\pi(\bar{x}^{(0)}_k-\bar{x}^{(0)}_s)+n\a=(n-2)\a^{(1)}-
nl\coth\pi(\bar{x}^{(0)}_s)\no\\
&&~~~~~~~~~~~~~~~~~~~~~~~~~~~~~~~~~~~~~+\sum_{k=1}^{N_2}
\coth\pi(\bar{x}^{(1)}_k-\bar{x}^{(0)}_s),\\
&&2\sum_{k=1,k\ne
s}^{N_{i+1}}\coth\pi(\bar{x}^{(i)}_k-\bar{x}^{(i)}_s)+(n-i)\a^{(i)}=
(n-i-2)\a^{(i+1)}-
\sum_{k=1}^{N_i}\coth\pi(\bar{x}_s^{(i)}-\bar{x}^{(i-1)}_k)\no\\
&&~~~~~~~~~~~~~~~~~~~~~~~~~~~~~~~~~~~~~+\sum_{k=1}^{N_{i+2}}
\coth\pi(\bar{x}^{(i+1)}_k-\bar{x}^{(i)}_s),~i=1,\cdots, n-3,\label{BA-C5}\\
&&2\sum_{k=1,k\ne
s}^{N_{n-1}}\coth\pi(\bar{x}^{(n-2)}_k-\bar{x}^{(n-2)}_s)+2\a^{(n-2)}
=-\sum_{k=1}^{N_{n-2}}\coth\pi(\bar{x}^{(n-2)}_s-\bar{x}^{(n-3)}_k).
\label{BA-C6}\eea Here the parameters $\a^{(i)},~i=0,1,\cdots,
n-2$ and $\a^{(0)}=\a$ are given by the relation (\ref{Relation})
from $n-1$ arbitrary non-negative integers
$\{\alpha^{(i)}\in\Zb^{+}|i=1,\cdots,n-1\}$.
\end{Proposition}

\vspace{0.6truecm}

\subsection{Rational potential}
Taking further rational limit of the elliptic Hamiltonian
(\ref{Ham-CM}) as in subsection 3.3, we can obtain the Hamiltonian
of rational CM model associated with $A_{n-1}$ root system \bea
H_{CM}=-\sum_{i=1}^n
\frac{\partial^2}{(\partial\l_i)^2}+\sum_{i\ne
j}\frac{l(l+1)}{(\l_i-\l_j)^2}. \label{Ham-Ra}\eea Moreover, we
have

\begin{Proposition}
The eigenvalues of the Hamiltonian (\ref{Ham-Ra}) of the rational
CM model associated with  $A_{n-1}$ root system with the discrete
coupling constants $\g=l+1$  are \bea
\e_{CM}=\sum_{k=1}^{N_1}\frac{(nl-1)n}{(\bar{x}^{(0)}_k)^2},\eea
where the $\frac{n(n-1)}{2}l$ real parameters
$\{\{\bar{x}^{(i)}_k\}\}$ satisfy the Bethe ansatz equations \bea
&&2\sum_{k=1,k\ne
s}^{N_1}\frac{1}{\bar{x}^{(0)}_k-\bar{x}^{(0)}_s}+n\a=(n-2)\a^{(1)}-
\frac{nl}{\bar{x}^{(0)}_s}+\sum_{k=1}^{N_2}
\frac{1}{\bar{x}^{(1)}_k-\bar{x}^{(0)}_s},\\
&&2\sum_{k=1,k\ne
s}^{N_{i+1}}\frac{1}{\bar{x}^{(i)}_k-\bar{x}^{(i)}_s}+(n-i)\a^{(i)}=
(n-i-2)\a^{(i+1)}-
\sum_{k=1}^{N_i}\frac{1}{\bar{x}_s^{(i)}-\bar{x}^{(i-1)}_k}\no\\
&&~~~~~~~~~~~~~~~~~~~~~~~~~~~~~~~+\sum_{k=1}^{N_{i+2}}
\frac{1}{\bar{x}^{(i+1)}_k-\bar{x}^{(i)}_s},~~i=1,\cdots, n-3,\label{BA-C7}\\
&&2\sum_{k=1,k\ne
s}^{N_{n-1}}\frac{1}{\bar{x}^{(n-2)}_k-\bar{x}^{(n-2)}_s}+2\a^{(n-2)}
=-\sum_{k=1}^{N_{n-2}}\frac{1}{\bar{x}^{(n-2)}_s-\bar{x}^{(n-3)}_k}.
\label{BA-C8}\eea Here the parameters $\a^{(i)},~i=0,1,\cdots,
n-2$ and $\a^{(0)}=\a$ are given by the relation (\ref{Relation})
from $n-1$ arbitrary non-negative real numbers
$\{\alpha^{(i)}\in\R^{+}|i=1,\cdots,n-1\}$.
\end{Proposition}

\vspace{0.6truecm}

\section{Summary and comments}
\label{Con} \setcounter{equation}{0}

Using the nested Bethe ansatz method for $\E$ \cite{Hou03}, we
obtain the spectrum of the Hamiltonian of all types of (elliptic,
trigonometric, rational) RS models associated with $A_{n-1}$ root
system with the discrete {\it coupling constant\/} $\g=\w l$.
Eigenvalues are given in the Bethe ansatz formulas (or the {\it
third formulas} in sense of Felder et al \cite{Fel95}). The
corresponding eigenfunction is a meromorphic  function of
$\{\l_i\}$ and has quasi-periodic properties (\ref{Qasi-P}) for
the elliptic and trigonometric cases, asymptotic properties
(\ref{asym}) for the rational case. For the special case of $n=2$,
our generalized result recovers that of \cite{Fel96,Rui98,Hik98}.

Taking the ``{\it non-relativistic limit\/}", the Hamiltonian of
RS model becomes that of the CM model. Then, we give eigenvalues
of the Hamiltonian of  CM models associated with  $A_{n-1}$ root
system with the discrete {\it coupling constant\/} $\g=l+1$ in the
Bethe ansatz formulas. Our formulas coincide with those of
\cite{Fel95} and those of $n=2$ case \cite{Rui98,Che99}. The
eigenvalues from our formulas are real and positive up to the {\it
ground state energy\/} $\e_0$ as physically desired. But, we have
not yet got a direct proof of positivity of the eigenvalues of RS
models from our formulas. However, we can show that for small {\it
coupling constant\/} the eigenvalues of RS model associated with
$A_{n-1}$ root system are positive, from their asymptotic
expansion (\ref{PHam}). Moreover, we find that the elliptic and
trigonometric RS and CM models  have  {\it discrete spectrum\/}
which are parameterized by a set of non-negative  integers
$\{\alpha^{(i)}|i=1,\cdots,n-1\}$. The rational RS and CM models
have {\it continuous spectrum\/} which are parameterized by a set
of non-negative real numbers $\{\alpha^{(i)}|i=1,\cdots,n-1\}$.

If one writes the Hamiltonian of  CM model  with the {\it coupling
constant} $\g$ as $H_{CM}(\g)$ and the associated eigenvalues as
$\e_{CM}(\g)$, from the expression of the Hamiltonian
(\ref{Ham-CM}), (\ref{Ham-Tr}) and (\ref{Ham-Ra}) one can find the
following {\it duality\/} \bea H_{CM}(-\g)=H_{CM}(\g-1).\eea Then,
actually, we have already obtained the spectrum of CM models
associated with  $A_{n-1}$ root system with the  discrete {\it
coupling constants\/} $\g=l~(l\in \Zb)$. Unfortunately, such a
{\it duality} does not exist for RS models associated with
$A_{n-1}$ root systems.

There also exists  another way (we call {\it symmetric polynomials
approach\/}) to get eigenfunctions and the corresponding
eigenvalues of the {\it trigonometric and rational} RS models
\cite{Mac95} and CM models \cite{Awa95,Lap96,Bak97,Kha00}. It
would be very interesting to compare our formulas (of
trigonometric and rational cases ) with those obtained by the {\it
symmetric polynomials approach\/} (for special $A_1$ case, it has
already been obtained \cite{Rui98}). However, the {\it symmetric
polynomials approach\/} fails in the elliptic models.

\section*{Acknowledgements}
We thank A. Belavin and T. Inami for  useful discussion. W.-L.
Yang is supported by the Japan Society for the Promotion of
Science.

\section*{Appendix A: Formulas for elliptic functions}
\setcounter{equation}{0}
\renewcommand{\theequation}{A.\arabic{equation}}
In this appendix, we give some useful series expansions of the
elliptic functions given by (\ref{Function}) when
$\tau=\sqrt{-1}\k,~\k\in\R,~\k>0$. By (\ref{Theta}), $\s$-function
can be expressed in terms of product form \cite{Bax86} (see
Chapter 15) \bea \s(u)=q^{\frac{1}{4}}\sin\pi u~
\prod_{n=1}^{\infty}(1-q^{2n}e^{2\sqrt{-1}\pi u})
(1-q^{2n}e^{-2\sqrt{-1}\pi
u})(1-q^{2n}),~~q=e^{-\pi\k}.\label{A1}\eea We can derive the
following series expansions from (\ref{Function}) \bea
&&\z(u)=\frac{\pi\cos\pi u}{\sin\pi u}+\pi\sum_{n=1}^{\infty}
\frac{\sin2\pi
u}{\sin\pi(u+\sqrt{-1}n\k)~\sin\pi(u-\sqrt{-1}n\k)},\label{A2}\\
&&\p(u)=\frac{\pi^2}{\sin^2\pi u}+\sum_{n=1}^{\infty}\lt\{
\frac{\pi^2}{\sin^2\pi(u+\sqrt{-1}n\k)}+
\frac{\pi^2}{\sin^2\pi(u-\sqrt{-1}n\k)}\rt\}. \label{A3}\eea
Moreover, the functions have following properties\bea
&&\p(\sqrt{-1}u)=-\lt\{\frac{\pi^2}{\sinh^2\pi
u}+\sum_{n=1}^{\infty}\lt\{ \frac{\pi^2}{\sinh^2\pi(u+n\k)}+
\frac{\pi^2}{\sinh^2\pi(u-n\k)}\rt\}\rt\},\label{Post}\\
&&\s^*(u)=\s(u^*),~\z^*(u)=\z(u^*),~\p^*(u)=\p(u^*),\label{real}
\eea where $*$ stands for the complex conjugation.


\end{document}